\documentclass[a4paper,12pt]{article}

\usepackage{amsmath,amsthm,amsbsy,amssymb,amsfonts,graphicx,mathrsfs,bm,accents,cite,xcolor,color}

\makeatletter
\catcode`\@=11
\@addtoreset{equation}{section}

\makeatother

\renewcommand{\thefootnote}{\arabic{footnote}}

\newcommand{\Exp}[1]{\operatorname{e}^{#1}}
\newcommand{\abs}[1]{\lvert {#1} \rvert}
\newcommand{\bra}[1]{\langle {#1} \rvert}
\newcommand{\ket}[1]{\lvert {#1} \rangle}
\newcommand{\rmd}{{\mathrm{d}}}
\newcommand{\nn}{\nonumber}
\newcommand{\Lie}{\pounds}
\newcommand{\gLie}{\hat{\pounds}}

\newcommand{\cA}{\mathcal A}
\newcommand{\cC}{\mathcal C}\newcommand{\cD}{\mathcal D}
\newcommand{\cE}{\mathcal E}\newcommand{\cF}{\mathcal F}
\newcommand{\cH}{\mathcal H}

\newcommand{\cK}{\mathcal K}\newcommand{\cL}{\mathcal L}

\newcommand{\cP}{\mathcal P}

\newcommand{\cS}{\mathcal S}

\newcommand{\sfa}{\mathsf{a}}
\newcommand{\sfi}{\mathsf{i}}
\newcommand{\sfm}{\mathsf{m}}

\newcommand{\sfx}{\mathsf{x}}
\newcommand{\sfy}{\mathsf{y}}
\newcommand{\sfA}{\mathsf{A}}
\newcommand{\sfC}{\mathsf{C}}
\newcommand{\sfF}{\mathsf{F}}
\newcommand{\sfM}{\mathsf{M}}

\newcommand{\rmT}{\mathrm{T}}
\newcommand{\OO}{\mathrm{O}}
\newcommand{\diag}{\mathrm{diag}}

\newcommand{\bnabla}{{\mathring{\nabla}}{}}
\newcommand{\bGamma}{{\mathring{\Gamma}}{}}
\newcommand{\bS}{{\mathring{\cS}}{}}
\newcommand{\bX}{{\bm X}}
\newcommand{\bY}{{\bm Y}}
\newcommand{\bB}{{\bm B}}
\newcommand{\bbS}{{\mathbb{S}}}
\newcommand{\bbX}{\mathbb{X}}

\newcommand{\bbeta}{{\bm\beta}}
\newcommand{\OG}{\tilde{g}}
\newcommand{\CG}{G}

\newcommand{\sla}[1]{\setbox0=\hbox{$#1$} 
\dimen0=\wd0 \setbox1=\hbox{/} \dimen1=\wd1 
\ifdim\dimen0>\dimen1 \rlap{\hbox to \dimen0{\hfil/\hfil}} #1 
\else\rlap{\hbox to \dimen1{\hfil$#1$\hfil}} / \fi}
\allowdisplaybreaks[3]

\begin{document}

\begin{titlepage}
\renewcommand{\thefootnote}{\fnsymbol{footnote}}

\begin{flushright}
\parbox{4cm}
{KUNS-2668}
\end{flushright}

\vspace*{1cm}

\centerline{{\Large\textbf{Weyl invariance for generalized supergravity backgrounds}}}
\vspace{2mm}
\centerline{\Large\textbf{from the doubled formalism}}%

\vspace{1.0cm}

\centerline{
{\large Jun-ichi Sakamoto$^a$},%
\footnote{E-mail address: \texttt{sakajun@gauge.scphys.kyoto-u.ac.jp}}
{\large Yuho Sakatani$^{b,c}$}%
\footnote{E-mail address: \texttt{yuho@koto.kpu-m.ac.jp}}
and {\large Kentaroh Yoshida$^a$}%
\footnote{E-mail address: \texttt{kyoshida@gauge.scphys.kyoto-u.ac.jp}}
}

\vspace{0.2cm}

\begin{center}
${}^a${\it Department of Physics, Kyoto University,}\\
{\it Kitashirakawa Oiwake-cho, Kyoto 606-8502, \rm Japan}

\vspace*{1mm}

${}^b${\it Department of Physics, Kyoto Prefectural University of Medicine,}\\
{\it Kyoto 606-0823, \rm Japan}

\vspace*{1mm}

${}^c${\it Fields, Gravity \& Strings, CTPU}\\
{\it Institute for Basic Sciences, Daejeon 34047 \rm Korea}
\end{center}

\vspace*{1mm}

\begin{abstract}
It has recently been shown that a set of the generalized type IIB supergravity equations 
follows from the requirement of kappa symmetry of the type IIB Green--Schwarz 
superstring theory defined on an arbitrary background. In this paper, we show that 
the whole bosonic part of the generalized type II supergravity equations can be reproduced 
from the $T$-duality covariant equations of motion of the double field theory 
by choosing a non-standard solution of the strong constraint. 
Then, by using the doubled formalism, we show the Weyl invariance of 
the bosonic string sigma model on a generalized gravity background. 
According to the dual-coordinate dependence of the dilaton, 
the Fradkin--Tseytlin term nicely removes the Weyl anomaly. 
This result seems likely to support that string theories can be consistently defined 
on arbitrary generalized supergravity backgrounds. 
\end{abstract}

\thispagestyle{empty}
\end{titlepage}

\setcounter{footnote}{0}

\tableofcontents

\section{Introduction}

A generalization of the type IIB supergravity recently proposed in \cite{Arutyunov:2015mqj} 
is a fascinating subject in string theory.%
\footnote{This was originally proposed to support a $q$-deformed AdS$_5\times$S$^5$ background 
\cite{Delduc:2013fga,Arutyunov:2013ega,Arutyunov:2015qva} as a solution.} 
This generalized system includes extra vector fields as well as the standard component fields 
of the type IIB supergravity. The classical action has not been revealed yet, 
and only the equations of motion are presented. 
Hereafter, we will refer to them as the generalized supergravity equations of motion (GSE) 
for simplicity. 

\medskip 

It is well known that the on-shell condition of type IIB supergravity ensures the 
kappa invariance of the Green--Schwarz string theory \cite{Grisaru:1985fv,Bergshoeff:1985su}. 
Conversely, in a recent paper \cite{WT}, the GSE have been reproduced 
by solving the kappa-symmetry constraints, generalizing the well-known fact. 
So far, it is considered that type IIB string theories on generalized supergravity 
backgrounds would not be Weyl invariant though still scale invariant. 
Therefore, it has not been clear whether the string theory is consistently 
defined on such backgrounds. 

\medskip 

It is worth noting that a $T$-duality transformation rule from 
a solution of the GSE to a solution of standard supergravity is given in \cite{Arutyunov:2015mqj}. 
On the other hand, one can map a solution of standard supergravity 
with a linear dilaton to a solution of the GSE by performing a formal $T$-duality 
transformation along a direction for which the dilaton is not isometric \cite{Hoare:2015wia}. 
These results indicate that solutions of standard supergravity and the GSE 
should be treated on an equal footing in the context of string theory, 
because the $T$-duality is a symmetry of string theory. 
However, there is a puzzle for the Weyl invariance. 
The Weyl invariance of string theories defined on solutions of the GSE may be broken, 
as discussed in \cite{Arutyunov:2015mqj}, 
though string theories on solutions of standard supergravity are Weyl invariant. 
Due to the $T$-duality symmetry, it seems likely that 
Weyl invariance should be preserved even for the solutions of the GSE in a certain manner.%
\footnote{We are grateful to A.~A.~Tseytlin for useful discussions on this point.} 

\medskip 

In order to clarify the issue of Weyl invariance under $T$-duality transformations, it is useful to utilize the manifestly $T$-duality-covariant 
formulations of supergravity and string theory: double field theory (DFT) \cite{Siegel:1993xq,Siegel:1993th,Siegel:1993bj,Hull:2009mi,Hull:2009zb,Hohm:2010jy,Hohm:2010pp}%
\footnote{For earlier observations along this direction, see footnotes 1 and 23 of \cite{Hoare:2015wia}.} 
and double sigma model (DSM) 
\cite{Tseytlin:1990nb,Tseytlin:1990va,Hull:2004in,Hull:2006va,Copland:2011wx,Lee:2013hma}. 
In the previous work \cite{Sakatani:2016fvh}, a modification of the DFT was studied so as to incorporate the GSE 
(in the context of the DFT for the NS--NS sector).%
\footnote{More recently, it was proposed in \cite{Baguet:2016prz} that the whole bosonic sector of the 
GSE can be reproduced from the manifestly $U$-duality-covariant formulation of supergravity, the 
exceptional field theory, by choosing a non-standard section and considering a certain Scherk--Schwarz-type ansatz. See Sect.~\ref{sec:discussion} for more details.} 
The modified DFT (mDFT) allows us to reproduce the GSE 
by choosing a section under which all of the fields do not depend on the dual coordinates.
In this paper, we will show that the extra generalized vector $\bX^M$ of the mDFT can always be removed with a redefinition of the dilaton. 
In this sense, the mDFT is no more than the usual DFT, and it should rather be called the (m)DFT. 
The point is that this redefinition introduces the dual-coordinate dependence into the dilaton (while keeping the strong constraint intact), 
and this dual-coordinate dependence is the origin of the modification of the supergravity equations of motion. 
Moreover, we extend the (m)DFT by including the R--R fields and show that the GSE are definitely 
reproduced from the $\OO(D,D)$ covariant equations of motion of the (m)DFT. 
We then find the relation between the R--R potentials and their strengths, the $\OO(D,D)$ transformation 
rule, and the generalized type IIA supergravity equations of motion. 
We also argue that the usual DFT action is nothing but the action for the GSE. 

\medskip 

After formulating the GSE from the perspective of the DFT, we will consider the string sigma model 
defined on the doubled target space, namely the DSM. 
As is well known, the Weyl anomaly of the string sigma model is canceled if the NS--NS background 
satisfies the supergravity equations of motion \cite{Callan:1985ia}. 
Here, the Weyl anomaly coming from the background metric $\CG_{mn}$ and the Kalb--Ramond field 
$B_{mn}$ is canceled by adding a counterterm to the string action, the Fradkin--Tseytlin term 
\cite{Fradkin:1983xs}. 
For a more general case in which the background satisfies the GSE, we can no longer find 
an appropriate counterterm to cancel the Weyl anomaly in the usual consideration, 
and hence the Weyl symmetry is broken to the scale symmetry \cite{Arutyunov:2015mqj} 
(see also \cite{Hull:1985rc,Tseytlin:1986tt,Shore:1986hk,Tseytlin:1986ws}). 
This has been the common understanding so far. 
In this paper, we consider the DSM defined on a general solution of the GSE, 
and elucidate that the Weyl anomaly can always be canceled 
by introducing a linear dual-coordinate dependence into the dilaton 
of the Fradkin--Tseytlin term.%
\footnote{For an earlier argument on the recovery of Weyl invariance in the doubled formalism, 
see, for example, footnote 23 of \cite{Arutyunov:2015mqj}.} 
In this sense, the usual supergravity backgrounds 
and solutions of the GSE can be treated on an equal footing in string theory. 

\medskip 

This paper is organized as follows. 
In Sect.~\ref{sec:(m)DFT}, after giving a short review of the mDFT \cite{Sakatani:2016fvh}, 
we reinterpret the mDFT as the usual DFT 
with a modified section. In Sect.~\ref{sec:R-R(m)DFT}, we review the R--R sector of the DFT and then 
reproduce the generalized type IIA and IIB supergravity equations by employing the modified section. 
We also present the $T$-duality transformation rule for solutions of the generalized type II supergravities. 
In Sect.~\ref{sec:Weyl}, we discuss the Weyl invariance of the string sigma model defined on a solution of the GSE. 
Section \ref{sec:discussion} is devoted to conclusions and discussion.

\section{NS--NS sector of (m)DFT}
\label{sec:(m)DFT}

In this section, we introduce the mDFT proposed in \cite{Sakatani:2016fvh}, 
in which only the NS--NS sector was studied and the R--R fields have not been included yet. 
We first give a short introduction to the mDFT. 
Then, we show that the mDFT can be regarded as the conventional DFT 
with a non-standard solution of the strong constraint. 

\subsection{A brief review of mDFT}

Let us give a short introduction to the mDFT.
In the absence of the R--R fields, the set of GSE in $D$ dimensions takes the following form: 
\begin{align}
\begin{split}
 &R_{mn} - \frac{1}{4}\,H_{mpq}\,H_n{}^{pq} + D_m X_n + D_n X_m = 0 \,,
\\
 &\frac{1}{2}\,D^k H_{kmn} - \bigl(X^k H_{kmn} + D_m X_n - D_n X_m \bigr) = 0 \,,
\\
 &R - \frac{1}{2}\, \abs{H_3}^2 + 4\,D_m X^m - 4\,X^m X_m = 0 \,,\qquad 
 X_m\equiv I_m +Z_m\,. 
\end{split}
\label{eq:GSE-NSNS}
\end{align}
Here we have defined $\abs{\alpha_p}^2\equiv \frac{1}{p!}\,
\alpha_{m_1\cdots m_p}\,\alpha^{m_1\cdots m_p}$, and $D$-dimensional indices $m,n,\cdots$ 
are raised or lowered with the metric $\CG_{mn}$. 
The covariant derivative $D_m$ is the conventional Levi--Civita connection associated 
with $\CG_{mn}$, and $H_{kmn}\equiv 3\,\partial_{[k}B_{mn]}$\,. 
A vector field $I^m$ and a 1-form $Z_m$ are defined so as to satisfy
\begin{align}
 D_m I_n + D_n I_m =0\,, \qquad 
 I^k\, H_{kmn} + D_m Z_n - D_n Z_m =0\,,\qquad I^m\,Z_m = 0\,. 
\label{eq:GSE-conditions}
\end{align}
The conventional dilaton is included in $Z_m$ as follows:
\begin{align}
 Z_m = \partial_m \Phi + U_m \,. 
\end{align}
Note that the equations of motion in \eqref{eq:GSE-NSNS} reduce to the conventional supergravity ones if $I^m=0$ and $U_m=0$ are satisfied. 
Since the GSE depend on $\Phi$ and $U_m$ only through the combination $Z_m$, there is an ambiguity in the decomposition of $Z=\rmd\Phi+U$ into $\rmd\Phi$ and $U$. 
Namely, at the level of the equations of motion, there is a local symmetry,
\begin{align}
 \Phi(x)\to \Phi(x) + \omega(x)\,,\qquad U(x)\to U(x)-\rmd \omega(x) \,.
\end{align}
Therefore, for a given solution of the GSE, we can always choose the dilaton to satisfy
\begin{align}
 \Lie_I \Phi = I^m\,\partial_m \Phi = 0 \,. 
\label{eq:dilaton-isometry}
\end{align}

\medskip 

In \cite{Sakatani:2016fvh}, by using techniques developed in the DFT, 
the above equations of motion have been reformulated 
in a manifestly $\OO(D,D)$ $T$-duality-covariant form,
\begin{align}
 \bS_{MN} = 0 \,,\qquad \bS = 0 \,,\qquad 
 \gLie_\bX \cH_{MN}=0\,,\qquad \gLie_\bX d=0\,,\qquad \bX^M \bX_M =0\,. 
\label{eq:mDFT-eom}
\end{align}
In order to explain these equations, let us begin with some basics 
(see \cite{Sakatani:2016fvh} for more details). 
We consider the equations of motion in a $2D$-dimensional doubled spacetime 
with the local coordinates $(x^M)=(x^m,\,\tilde{x}_m)$, 
where $\tilde{x}_m$ are called the dual coordinates 
while $x^m$ are the conventional coordinates in the supergravity. 
We then introduce the generalized metric $\cH_{MN}$ on the doubled spacetime, 
which can be parameterized as
\begin{align}
 \cH(x)=\bigl(\cH_{MN}\bigr) = \begin{pmatrix} \CG_{mn}-B_{mp}\,\CG^{pq}\,B_{qn} & B_{mk}\,\CG^{kn} \\
 -\CG^{mk}\,B_{kn} & \CG^{mn} 
 \end{pmatrix}
\label{eq:H_MN-conventional}
\end{align}
in terms of the conventional metric $\CG_{mn}$ and the Kalb--Ramond field $B_{mn}$\,. 
The $T$-duality-invariant dilaton $d(x)$, often called the DFT dilaton, 
can be related to the conventional dilaton $\Phi(x)$ as
\begin{align}
 \Exp{-2d}= \Exp{-2\Phi}\sqrt{\abs{\CG}}\,. 
\label{eq:dilaton-def}
\end{align}
A generalized vector field $\bX^M$, which is absent in the conventional DFT, is parameterized as
\begin{align}
 \bigl(\bX^M\bigr) = \begin{pmatrix} I^m \\ U_m + B_{mn}\, I^n
 \end{pmatrix} \,,
\label{eq:X-parameterization}
\end{align}
where $I^m$ and $U_n$ here are identified with the ones appearing in the GSE. 
The $2D$-dimensional indices $M,N,\cdots$ are raised or lowered with the $\OO(D,D)$ metric,
\begin{align}
 (\eta_{MN})\equiv \begin{pmatrix} 0 & \delta_m^n \\ \delta^m_n & 0 \end{pmatrix} \,,\qquad 
 (\eta^{MN}) \equiv \begin{pmatrix} 0 & \delta^m_n \\ \delta_m^n & 0 \end{pmatrix} \,. 
\end{align}
The generalized diffeomorphisms in the doubled spacetime are generated 
by the generalized Lie derivative $\gLie_V$, which acts on $\cH_{MN}(x)$ and $d(x)$ as
\begin{align}
\begin{split}
 \gLie_V \cH_{MN} &= V^K\,\partial_K \cH_{MN} 
 + \big(\partial_M V^K -\partial^K V_M\big)\, \cH_{KN} 
 + \big(\partial_N V^K -\partial^K V_N\big)\, \cH_{MK} \,, 
\\
 \gLie_V \Exp{-2d} &= \partial_M \bigl(\Exp{-2d}V^M\bigr) \,. 
\end{split}
\end{align}
We suppose that all of the fields and gauge parameters satisfy the so-called strong constraint,
\begin{align}
\label{strong}
 \eta^{MN}\,\partial_M A(x)\,\partial_N B(x) 
 = \partial_m A(x)\, \tilde{\partial}^m B(x) + \tilde{\partial}^m A(x)\, \partial_m B(x) = 0 \,,
\end{align}
where $A(x)$ and $B(x)$ are fields or gauge parameters. 
Then, the generalized diffeomorphisms can be regarded as gauge symmetries of the DFT, 
and the associated gauge algebra is closed. 
In order to satisfy the strong constraint \eqref{strong}, we usually consider a solution in which 
all of the fields and gauge parameters are independent of $\tilde{x}_m$\,. 
On the other hand, we will take a different solution when we describe the GSE, as we will discuss in Sect.~\ref{sec:mDFT-is-DFT}. 

\paragraph{Equations of motion:} 
Let us now explain the equations of motion \eqref{eq:mDFT-eom}. 
The last three equations in \eqref{eq:mDFT-eom}, which reproduce \eqref{eq:GSE-conditions} and \eqref{eq:dilaton-isometry}, indicate that the generalized vector $\bX^M$ is a null generalized Killing vector. 
On the other hand, the first two equations in \eqref{eq:mDFT-eom} describe 
the dynamics of $\cH_{MN}(x)$ and $d(x)$\,. 
In particular, the first equation reproduces the first two equations in \eqref{eq:GSE-NSNS} 
and the second equation leads to the last equation in \eqref{eq:GSE-NSNS}. 
In fact, $\bS_{MN}$ and $\bS$ are the generalized Ricci tensor/scalar associated 
with the covariant derivative satisfying (see \cite{Sakatani:2016fvh} for more details)
\begin{align}
 \bnabla_K \eta_{MN}=0\,,\qquad 
 \gLie_V=\gLie^{\nabla}_V\,,\qquad 
 \bnabla_K \cH_{MN}=0\,,\qquad 
 \bnabla_M d + \bX_M = 0 \,. 
\label{eq:generalized-connection}
\end{align}
The explicit expressions of the modified quantities, the generalized connection 
$\bGamma_{MNK}$, the generalized Ricci tensor $\bS_{MN}$, and the generalized Ricci scalar $\bS$, 
in terms of $(\cH_{MN},\, d,\, \bX^M)$ or $(\CG_{mn},\,B_{mn},\,\Phi,\,I^m,\,U_m)$, 
can be found in the preceding paper \cite{Sakatani:2016fvh} (see Sects.~3.2, 4.1, and 4.2 therein). 
The corresponding quantities in the conventional DFT, $\Gamma_{MNK}$, $\cS_{MN}$, 
and $\cS$, can be reproduced from these modified quantities by setting $\bX^M=0$. 
Conversely, the modified quantities $(\bGamma_{MNK},\,\bS_{MN},\,\bS)$ can be obtained 
from $(\Gamma_{MNK},\,\cS_{MN},\,\cS)$ with the replacement
\begin{align}
 \partial_M d\ \to\ \partial_M d +\bX_M \,. 
\label{eq:modification}
\end{align}
The meaning of this shift will be clarified in the next subsection. 

\subsection{(m)DFT for ``DFT on a modified section''}
\label{sec:mDFT-is-DFT}

In this subsection, we show that the mDFT, which was reviewed in the previous subsection, 
is equivalent to the conventional DFT with a non-standard solution of the strong constraint. 
In this sense, the mDFT should rather be called the (m)DFT. 

\medskip 

Let us first prove that by performing a certain generalized coordinate transformation, 
the null generalized Killing vector $\bX^M$ can always be brought into the following form:%
\footnote{Our proof partially follows the discussion given in Sect.~3.1 of \cite{Hull:2006qs}.}
\begin{align}
 \bX^M \equiv \begin{pmatrix} I^m \\ U_m + B_{mn}\, I^n
 \end{pmatrix} = \begin{pmatrix} I^m \\ 0
 \end{pmatrix}\qquad (I^m:\text{constant})\,. 
\label{eq:X-gauge}
\end{align}
This statement can be regarded as a generalization of the well-known fact 
in the Riemannian geometry that we can always find a certain coordinate system 
where the components of a Killing vector are constant (see, for example, \cite{Wald}). 

\medskip 

From the strong constraint, we can always find a section where all of the fields 
$(\cH_{MN},\,d,\,\bX^M)$ are independent of the dual coordinates. 
With this choice of section, the null and the generalized Killing properties 
lead to the conditions \eqref{eq:GSE-conditions} and \eqref{eq:dilaton-isometry}. 
Since $I^m$ is a Killing vector field, we can always find a certain coordinate system $(x^m)=(x^\mu,\,y)$ 
in which the Killing vector is a coordinate basis: $I^m= c\,\delta^m_y$\,, where $c$ is a constant. 
In such a coordinate system, both $\CG_{mn}$ and $\Phi$ are independent of $y$\,. 
The 3-form $H_3$ is also independent of $y$, as we can easily show $\Lie_I H_3=0$ from \eqref{eq:GSE-conditions}. 
Thus, we can generally expand $H_3$ as
\begin{align}
 H_3 = h_3 + c^{-1}\,\iota_I H_3\wedge \rmd y = h_3 - c^{-1}\,\rmd Z \wedge \rmd y \qquad \bigl(\iota_I h_3=0\bigr)\,,
\end{align}
where we used \eqref{eq:GSE-conditions}, and $h_3$ should satisfy $\Lie_I h_3=0$ 
that follows from $\Lie_I H_3=0$ and $\Lie_I Z=0$\,. 
From this expansion, we find an expansion of the $B$-field satisfying $H_3=\rmd B_2$\,,
\begin{align}
 B_2 = b_2 - c^{-1}\,U \wedge \rmd y \qquad \bigl(\iota_I b_2=0\,, \quad h_3=\rmd b_2 \bigr)\,,
\label{eq:B_2-solution}
\end{align}
where we used $\rmd Z=\rmd U$, and $b_2$ can always be chosen 
such that $\Lie_I b_2=0$ is satisfied. 
This shows that we can always take a gauge (for generalized diffeomorphisms) so that $B_{mn}$ is also independent of $y$ (i.e., $\Lie_I B_2=0$), 
and all of the NS--NS fields are now independent of $y$\,. 
From \eqref{eq:B_2-solution} and $\iota_I U=0$, which comes from \eqref{eq:GSE-conditions} and \eqref{eq:dilaton-isometry}, we also find the relation
\begin{align}
 \iota_I B_2 - U=0 \,.
\label{eq:i_IB=U}
\end{align}
This completes the proof that a null generalized Killing vector $\bX^M$ can always be brought into the form \eqref{eq:X-gauge}. 

\medskip 

Then, since all of the fields are independent of $y$\,, 
the $\tilde{y}$ dependence can sneak in without violating the strong constraint. 
Indeed, in a coordinate system where \eqref{eq:X-gauge} is realized, 
the shift \eqref{eq:modification} from the DFT to the mDFT can be interpreted 
as an implicit introduction of the linear $\tilde{y}$ dependence 
into the dilaton $d_*(x)$ or $\Phi_*(x)$:
\begin{align}
 (\partial_M d)=\begin{pmatrix}
 \partial_m d\\ 0
 \end{pmatrix}\quad \to\quad (\partial_M d+\bX_M)=\begin{pmatrix}
 \partial_m d \\ I^m
 \end{pmatrix} = 
 (\partial_M d_*) \,,\qquad 
 d_* \equiv d+ c\,\tilde{y} \,. 
\end{align}
When the dilaton does not depend on $\tilde{y}$ (i.e., $c=0$), 
$I^m$ vanishes and the modification disappears. 
From the above argument, solutions of the mDFT can always be described 
as solutions of the DFT in which the dilaton has a linear dual-coordinate dependence. 

\medskip 

Conversely, let us consider a solution of the DFT where the DFT dilaton has a linear dual-coordinate dependence, 
$d_*=d(x)+c^m \,\tilde{y}_m$ with constant $c^m$\,. 
From the identification
\begin{align}
 \partial_M d + \bX_M = \partial_M d_*\,,
\end{align}
we obtain that 
\begin{align}
 I^m=c^m\,, \qquad U_m + B_{mn}\, I^n=0\,.
\end{align}
Note here that $d_*$, say the (m)DFT dilaton, can have only linear dependence 
on the dual coordinates with constant coefficients if we prefer to avoid 
the explicit appearance of the dual coordinates in $I^m$ and $\partial_m d$\,. 
Suppose that all of the fields, collectively denoted by $\varphi$, are independent of $\tilde{y}$\,. 
Then the strong constraint requires that
\begin{align}
 0 = \partial^M d\,\partial_M \varphi = I^m \,\partial_m \varphi = \Lie_I \varphi \,. 
\end{align}
Here, in the last equality we have used the fact that $I^m$ is constant 
in order to express the condition in a covariant form. 
Then, the following conditions, namely \eqref{eq:GSE-conditions} and \eqref{eq:dilaton-isometry}, are automatically satisfied:
\begin{align}
 \Lie_I \CG_{mn} = 0\,, \quad I^m\,\partial_m\Phi = 0\,, \quad \iota_I H_3 + \rmd U =0 \,, 
\quad I^m\, U_m =0 \,. 
\label{eq:condition-section}
\end{align}
Here, the dilaton $\Phi$ is defined through the relation
\begin{align}
 \Exp{-2d}=\Exp{-2\Phi}\sqrt{\abs{\CG}}\,,
\end{align}
and it is independent of $\tilde{y}$\,. 
When the R--R fields are also introduced, they should also satisfy
\begin{align}
 \Lie_I \cF_p = 0\,. 
\end{align}

\medskip 

From the above viewpoint, it is not necessary to look for the action for the GSE. 
The DFT action supplies the $2D$-dimensional equations of motion of the DFT. 
If the DFT dilaton has the dual-coordinate dependence, the equations of motion take the form of the GSE. 

\section{Ramond--Ramond sector of (m)DFT}
\label{sec:R-R(m)DFT}

In this section, we introduce the R--R fields by following the well-established formulation of the DFT 
\cite{Hohm:2011zr,Hohm:2011dv,Geissbuhler:2011mx,Jeon:2012kd,Geissbuhler:2013uka}.%%%
\footnote{The approaches in \cite{Hohm:2011zr,Hohm:2011dv}, \cite{Jeon:2012kd}, 
and \cite{Geissbuhler:2011mx,Geissbuhler:2013uka}, respectively, are slightly different from each other. 
In this paper, we basically follow the approach of \cite{Hohm:2011zr,Hohm:2011dv}. 
It is also useful to follow \cite{Jeon:2012kd} when we consider the type II supersymmetric DFT 
\cite{Jeon:2012hp}.}
%%%
The whole bosonic part of the GSE is reproduced from the equations of motion of the DFT 
by choosing a modified section. 

\subsection{Ramond--Ramond sector of DFT}

\paragraph*{Gamma matrices and O$(D,D)$ spinors\\}

It is convenient to introduce the gamma matrices $\{\gamma^M\} = \{\gamma^m,\,\gamma_m\}$ 
satisfying the $\OO(D,D)$ Clifford algebra,%
\footnote{In this paper, we call the Pin$(D,D)$ group simply $\OO(D,D)$\,. }
\begin{align}
 \bigl\{\gamma^M,\, \gamma^N \bigr\} = \eta^{MN}\,. 
\end{align}
Here, the gamma matrices are real and satisfy $\gamma^m=(\gamma_m)^\rmT$\,. 
The gamma matrices with multi-indices are defined as 
\begin{align}
 \gamma^{M_1\cdots M_p}\equiv \gamma^{[M_1}\cdots \gamma^{M_p]}\,, \qquad 
 \gamma^{m_1\cdots m_p}\equiv \gamma^{[m_1}\cdots \gamma^{m_p]} = \gamma^{m_1}\cdots \gamma^{m_p}\,.
\end{align}
From the anti-commutation relations,
\begin{align}
 \{\gamma^m,\,\gamma_n\}=\delta^m_n\,,\qquad \{\gamma^m,\,\gamma^n\}=0
=\{\gamma_m,\,\gamma_n\} \,,
\end{align}
the $\gamma_m$ can be regarded as fermionic annihilation operators. 
The Clifford vacuum $\ket{0}$ is defined so as to satisfy
\begin{align}
 \gamma_m\ket{0}=0\,,\qquad \langle 0\vert 0\rangle = 1 \,,
\end{align}
where $\bra{0}\equiv \ket{0}^\rmT$. 
By acting $\gamma^m$ matrices on $\ket{0}$, an $\OO(D,D)$ spinor can be constructed as 
\begin{align}
 \ket{\alpha_p} \equiv \frac{1}{p!}\,\alpha_{m_1\dots m_p}\,\gamma^{m_1\cdots m_p}\ket{0}\,,
\end{align}
and it is in one-to-one correspondence with a $p$-form,
$\alpha_p \equiv\frac{1}{p!}\,\alpha_{m_1\dots m_p}\,\rmd x^{m_1}\wedge\cdots\wedge \rmd x^{m_p}$\,. 
A formal sum of $\OO(D,D)$ spinors,
\begin{align}
 \ket{\alpha}\equiv \sum_p \frac{1}{p!}\,\alpha_{m_1\dots m_p}\,\gamma^{m_1\cdots m_p}\ket{0}\,,
\end{align}
corresponds to a poly-form $\alpha=\sum_p \alpha_p$\,. 

\medskip

The $\OO(D,D)$ transformations are generated by $\gamma^{MN}$ satisfying 
\begin{align}
 [\gamma^{MN},\,\gamma_L] = \gamma_K\,(T^{MN})^K{}_L\,,\qquad 
 (T^{MN})^K{}_L \equiv 2\,\eta^{K[M}\,\delta^{N]}_L \,. 
\end{align}
By utilizing the generators, one can define the following quantities: 
\begin{align}
 S_{e^\rmT} \equiv \Exp{\frac{1}{2}\,h^m{}_n\,[\gamma_m,\,\gamma^n]} \quad 
\bigl[e_m{}^n\equiv (\Exp{h^\rmT})_m{}^n\bigr]\,,\qquad
 \Exp{\bB} \equiv \Exp{\frac{1}{2}\,B_{mn}\,\gamma^{mn}} \,,\qquad
 \Exp{\bbeta} \equiv \Exp{\frac{1}{2}\,\beta^{mn}\,\gamma_{mn}} \,. 
\end{align}
We can easily show that they satisfy
\begin{align}
\begin{split}
 S_{e^\rmT} \,\gamma_N S_e^{-1} &= \gamma_M\,(\Lambda_{e^\rmT})^M{}_N\,,\quad
 \Exp{\bB}\, \gamma_N \Exp{-\bB} = \gamma_M\,(\Lambda_B)^M{}_N \,,\quad
 \Exp{\bbeta}\, \gamma_N \Exp{-\bbeta} = \gamma_M\,(\Lambda_\beta)^M{}_N \,,
\\
 \Lambda_{e^\rmT}&\equiv\begin{pmatrix}
 (e^\rmT)^m{}_n & 0 \\ 0 & (e^{-1})_m{}^n
 \end{pmatrix} \,,\quad
 \Lambda_B\equiv\begin{pmatrix}
 \delta^m_n & 0 \\ B_{mn} & \delta_m^n
 \end{pmatrix} \,,\quad
 \Lambda_\beta\equiv\begin{pmatrix}
 \delta^m_n & \beta^{mn} \\ 0 & \delta_m^n
 \end{pmatrix} \,.
\end{split}
\end{align}
It is helpful to define the correspondent of the flat metric as 
\begin{align}
\begin{split}
 &S_k\equiv \gamma_0\,\gamma^0 - \gamma^0\,\gamma_0 = S_k^{-1}= S_k^\rmT \,,
\qquad 
 S_k\,\gamma_N\,S_k^{-1} = \gamma_M\,k^M{}_N \,,
\\
 &(k^M{}_N)\equiv \begin{pmatrix}
 k^m{}_n & 0 \\ 0 & k_m{}^n
 \end{pmatrix} \,, \quad 
 (k^m{}_n) \equiv \diag(-1,+1,\dotsc,+1) \equiv (k_m{}^n) \equiv (k_{mn}) \,. 
\end{split}
\end{align}
The correspondent of the $B$-untwisted metric,%
\footnote{See \cite{Hull:2014mxa} for discussions of the untwisted form of generalized tensors.}
\begin{align}
 (\hat{\cH}_{MN})\equiv \begin{pmatrix} \CG_{mn} & 0 \\ 0 & \CG^{mn} \end{pmatrix}\,,\qquad 
 \CG_{mn}\equiv e_m{}^k\,e_n{}^l\,k_{kl}\,,\qquad (\CG^{mn})\equiv (\CG_{mn})^{-1}\,,
\end{align}
can also be defined as
\begin{align}
 S_{\hat{\cH}} \equiv S_e\,S_k\,S_{e^\rmT} =S_{\hat{\cH}}^\rmT\,,\quad 
S_e\equiv (S_{e^\rmT})^\rmT\,, \quad 
 S_{\hat{\cH}} \,\gamma_N\,S_{\hat{\cH}}^{-1} = (\gamma^M)^\rmT\,\hat{\cH}_{MN}\,. 
\end{align}
This gives a natural metric,
\begin{align}
 \bra{\alpha}S_{\hat{\cH}} \ket{\beta} &= \sum_{p,q}\frac{1}{p!\,q!}\,
\alpha_{m_1\cdots m_p}\,\beta_{n_1\cdots n_q}\bra{0}\,\gamma_{m_p\cdots m_1} \,S_{\hat{\cH}} \, 
\gamma^{n_1\cdots n_q} \ket{0}
\nn\\
 &= \sum_{p,q}\frac{1}{p!\,q!}\,\alpha_{m_1\cdots m_p}\,\beta_{n_1\cdots n_q}\bra{0}\,
\gamma_{m_p\cdots m_1}\,(\gamma_{l_1\cdots l_q})^\rmT \,S_{\hat{\cH}}\, \ket{0}\,\CG^{l_1n_1}
\cdots \CG^{l_qn_q}
\nn\\
 &= \sqrt{\abs{\CG}}\sum_p \frac{1}{p!}\, \CG^{m_1n_1}\cdots 
\CG^{m_pn_p}\,\alpha_{m_1\cdots m_p}\,\beta_{n_1\cdots n_p} 
 = \bra{\beta}S_{\hat{\cH}}\ket{\alpha} \,.
\label{eq:inner-product}
\end{align}

\medskip

The charge conjugation matrix is defined as 
\begin{align}
\begin{split}
 &C \equiv (\gamma^0\pm \gamma_0) \cdots (\gamma^{D-1}\pm \gamma_{D-1}) 
\qquad (D:\text{even/odd})\,, 
\\
 &C\,\gamma^M\,C^{-1} = -(\gamma^M)^\rmT \,,\qquad C^{-1}= (-1)^{\frac{D(D+1)}{2}}C = C^\rmT \,.
\end{split}
\end{align}
By using $C$ and $S_{\hat{\cH}}$, the Hodge dual can be constructed as 
\begin{align}
 C\,S_{\hat{\cH}} \, \gamma^{m_1\cdots m_p}\ket{0} 
 &= \sqrt{\abs{\CG}}\, (-1)^p \CG^{m_1n_1}\cdots \CG^{m_pn_p}\, \gamma_{n_1\cdots n_p} \, C \ket{0} 
\nn\\
 &= \frac{1}{(D-p)!}\, (-1)^{\frac{p(p+1)}{2}}\, 
\varepsilon^{m_1\cdots m_p}{}_{n_1\cdots n_{D-p}} \, \gamma^{n_1\cdots n_{D-p}} \ket{0} \,, 
\label{eq:SH-Gamma}
\end{align}
where $\varepsilon_{01\cdots (D-1)}=+\sqrt{\abs{\CG}}$ and indices are raised or lowered 
with $\CG_{mn}$.

\medskip

The correspondent of the generalized metric is defined as 
\begin{align}
\begin{split}
 &S_\cH \equiv \Exp{-\bB^\rmT} S_{\hat{\cH}} \Exp{-\bB} = S_\cH^\rmT\,,\qquad 
 S_\cH \,\gamma_N\,S_\cH^{-1} = (\gamma^M)^\rmT\,\cH_{MN}\,,
\\
 &(\cH_{MN}) = \begin{pmatrix} \delta_m^k & B_{mk} \\
  0 & \delta^m_k 
 \end{pmatrix}\begin{pmatrix} \CG_{kl} & 0 \\
  0 & \CG^{kl} 
 \end{pmatrix}\begin{pmatrix} \delta^l_n & 0 \\
  -B_{ln} & \delta_l^n 
 \end{pmatrix}\,. 
\end{split}
\end{align}
As stressed in \cite{Hohm:2011dv}, $S_\cH$ is a particular parameterization of the fundamental field 
$\bbS$ that corresponds to the generalized metric $\cH_{MN}$ before providing a parameterization. 
If we take another parameterization of the generalized metric,
\begin{align}
 (\cH_{MN}) = \begin{pmatrix} \delta_m^k & 0 \\
  -\beta^{mk} & \delta^m_k 
 \end{pmatrix} \begin{pmatrix} \OG_{kl} & 0 \\
  0 & \OG^{kl} 
 \end{pmatrix} \begin{pmatrix} \delta^l_n & \beta^{kn} \\
  0 & \delta_l^n 
 \end{pmatrix} \,,
\label{eq:non-geometric-parameterization}
\end{align}
then the field $\bbS$ is parameterized as
\begin{align}
 S_{\tilde{\cH}} \equiv \Exp{\bbeta^\rmT} S_{\check{\cH}} \Exp{\bbeta}\,,\qquad 
 S_{\check{\cH}} \equiv S_{\tilde{e}}\,S_k\,S_{\tilde{e}^\rmT} \,.
\end{align}

\medskip

For later discussion, it is also convenient to define the following quantity,
\begin{align}
 \cK \equiv -C\,\bbS = C\,\cK^\rmT\,C \,, \qquad \cK\,\gamma_M\,\cK^{-1} 
= -\cH_M{}^N \, \gamma_N\,,
\end{align}
and the chirality operator
\begin{align}
 \gamma^{D+1} \equiv (-1)^{N_F}\,,\qquad N_F\equiv \gamma^m\,\gamma_m \,. 
\end{align}
The chirality operator acts on an $\OO(D,D)$ spinor as
\begin{align}
 \gamma^{D+1} \,\ket{\alpha} = \sum_p (-1)^p \ket{\alpha_p} \,.
\end{align}
A chiral/anti-chiral spinor corresponds to a poly-form with even/odd degree.

\paragraph*{The classical action and equations of motion\\}

The dynamical fields that correspond to the R--R fields are introduced as $\OO(D,D)$ spinors,
\begin{align}
 \ket{A} \equiv \sum_p \frac{1}{p!}\, A_{m_1\cdots m_p}\,\gamma^{m_1\cdots m_p}\ket{0} \,,
\end{align}
which transform under generalized diffeomorphisms \cite{Hohm:2011zr,Hohm:2011dv} as follows:
\begin{align}
 \delta_V \ket{A} = \gLie_V \ket{A}
 \equiv V^M\,\partial_M \ket{A} + \partial_M V_N\, \gamma^M\,\gamma^N \,\ket{A} \,.
\label{eq:generalized-diffeo-RR}
\end{align}
Depending on the type IIA or IIB theory, $\ket{A}$ takes a definite chirality,
\begin{align}
 \gamma^{11}\ket{A} = \mp \ket{A} \quad (\text{IIA/IIB})\,.
\label{eq:chirality}
\end{align}
It is easy to show that under the strong constraint, the R--R field strength,
\begin{align}
 \ket{F} \equiv \sla{\partial} \ket{A} \equiv \gamma^M\, \partial_M \ket{A} \,,
\label{eq:F-DA}
\end{align}
transforms covariantly under generalized diffeomorphisms. 

\medskip

From the strong constraint, the operator $\sla{\partial}$ is nilpotent, and one can readily see 
that the field strength is invariant under the gauge transformations for the R--R fields,
\begin{align}
 \delta_\lambda \ket{A} = \sla{\partial} \ket{\lambda}\,,
\end{align}
where $\ket{\lambda}$ is an arbitrary $\OO(D,D)$ spinor which respects the chirality \eqref{eq:chirality}. 
The Bianchi identity also follows from the nilpotency $\sla{\partial}^2=0$,
\begin{align}
 \sla{\partial} \ket{F} = \sla{\partial}^2 \ket{A} = 0 \,. 
\label{eq:Bianchi-id}
\end{align}

\medskip

In the democratic formulation \cite{Fukuma:1999jt,Bergshoeff:2001pv}, the self-duality relation should be imposed 
at the level of the equations of motion. 
In our convention, it takes the form
\begin{align}
 \ket{F} = \cK\, \ket{F} \,. 
\label{eq:self-duality}
\end{align}
From this expression and the Bianchi identity, we obtain the following relation:
\begin{align}
 \sla{\partial} \cK\, \ket{F} = 0 \,. 
\label{eq:eom-RR}
\end{align}
This is nothing but the equation of motion for the R--R field, as we will see below. 

\medskip

Now, let us write down the bosonic part of the (pseudo-)action for the type II supergravity,
\begin{align}
 \cL = \Exp{-2d} \cS -\frac{1}{4}\,\bra{F}\, \bbS\, \ket{F} 
 = \Exp{-2d} \cS + \frac{1}{4}\,\overline{\bra{F}}\, \cK\, \ket{F} \,,
\label{eq:Lagrangian-full}
\end{align}
where we have defined $\overline{\bra{F}}\equiv \bra{F}\,C^\rmT$. 
Taking a variation with respect to $A$, we obtain
\begin{align}
 \delta_A\cL = \frac{1}{2}\,\overline{\bra{\delta A}}\,\sla{\partial} \cK\, \ket{F}
 - \partial_M \Bigl[ \frac{1}{2}\,\overline{\bra{\delta A}}\, \gamma^M\,\cK\, \ket{F}\Bigr] \,,
\end{align}
and as expected, the equations of motion for $A$ reproduce \eqref{eq:eom-RR}. 
The variation with respect to the DFT dilaton becomes
\begin{align}
 \delta_d \cL = - 2\Exp{-2d} \cS\,\delta d 
 + \partial_M \bigl(4\Exp{-2d} \cH^{MN}\, \partial_N \delta d \bigr) \,, 
\end{align}
and there is no contribution from the R--R fields. 
Finally, the variation with respect to the generalized metric gives rise to 
\begin{align}
 \delta_\cH \cL &= - \frac{1}{2} \Exp{-2d} \cS^{MN}\, \delta \cH_{MN} 
 - \partial_M \bigl[\Exp{-2d} \nabla_N\,\delta \cH^{MN} \bigr]
 + \frac{1}{4}\, \overline{\bra{F}}\, \delta \cK\, \ket{F} 
\nn\\
 &= - \frac{1}{2} \Bigl(\Exp{-2d} \cS^{MN}- \frac{1}{4}\, \cH^{(M}{}_K\,\overline{\bra{F}}\,
\gamma^{N)K}\,\cK\, \ket{F}\Bigr)\, \delta \cH_{MN} 
 - \partial_M \bigl[\Exp{-2d} \nabla_N\,\delta \cH^{MN} \bigr] 
\nn\\
 &= - \frac{1}{2}\,\Exp{-2d} \bigl(\cS^{MN} + \cE^{MN}\bigr)\, \delta \cH_{MN} 
 - \partial_M \bigl[\Exp{-2d} \nabla_N\,\delta \cH^{MN} \bigr] \,,
\end{align}
where we have employed the identity \cite{Hohm:2011dv}
\begin{align}
 \delta \cK = \frac{1}{2}\,\cH^{(M}{}_K\,\gamma^{N)K}\,\cK\,\delta\cH_{MN} \,, 
\end{align} 
in the second equality, and defined the ``energy--momentum tensor'' \cite{Hohm:2011zr,Hohm:2011dv}
\begin{align}
 \cE^{MN} \equiv \frac{1}{4} \Exp{2d}\,\Bigl[\bra{F} \,(\gamma^{(M})^\rmT\,\bbS\,\gamma^{N)}\, 
\ket{F} - \frac{1}{2}\,\cH^{MN}\,\bra{F}\,\bbS\,\ket{F}\Bigr] \,.
\end{align}

\medskip

In summary, the equations of motion of the DFT are given by 
\begin{align}
 \cS_{MN} + \cE_{MN} = 0 \,,\qquad \cS = 0 \,, \qquad 
 \sla{\partial} \, \cK\, \ket{F} = 0\,.
\label{eq:EOM-DFT}
\end{align}

\paragraph*{The classical action and equations of motion in the conventional formulation\\}

Next, let us show that the expressions we obtained above indeed reproduce 
the well-known expressions in conventional supergravity 
by choosing \eqref{eq:H_MN-conventional}, \eqref{eq:dilaton-def}, and $\tilde{\partial}^m=0$. 

\medskip

From \eqref{eq:generalized-diffeo-RR} and $\tilde{\partial}^m=0$, 
the transformation of the R--R field under a generalized diffeomorphism becomes
\begin{align}
 \delta_V\ket{A}
 &= \bigl[v^m\,\partial_m + \partial_m v^n\,\gamma^m\,\gamma_n 
+ \tfrac{1}{2}\,(\partial_m\tilde{v}_n-\partial_n\tilde{v}_m)\,\gamma^{mn}\bigr]\,\ket{A}
\nn\\
 &= \ket{\Lie_v A + \rmd \tilde{v}\wedge A} \,,
\end{align}
and it is equivalent to a conventional diffeomorphism and a $B$-field gauge transformation. 
The field strength \eqref{eq:F-DA} and the Bianchi identity \eqref{eq:Bianchi-id} take the following form: 
\begin{align}
 \ket{\cF} = \sla{\partial} \ket{A} = \gamma^m\, \partial_m \ket{A} = \ket{\rmd A} \,, \qquad 
 \sla{\partial}\ket{\cF} = \ket{\rmd\cF} =0 \,. 
\end{align}
The self-duality relation for the R--R field strength becomes
\begin{align}
 \ket{F} = -C\,\Exp{-\bB^\rmT} S_{\hat{\cH}} \Exp{-\bB}\, 
\ket{F} = - \Exp{\bB} C\, S_{\hat{\cH}} \Exp{-\bB}\, \ket{F} \,. 
\label{eq:self-dual0}
\end{align}
If the $B$-untwisted field strength is defined as 
\begin{align}
 \ket{\hat{F}} \equiv \Exp{-\bB}\, \ket{F} \,,
\end{align}
which is invariant under the $B$-field gauge transformations, \eqref{eq:self-dual0} can be rewritten as
\begin{align}
 \ket{\hat{F}} = - C\, S_{\hat{\cH}} \, \ket{\hat{F}} \,. 
\end{align}
From \eqref{eq:SH-Gamma}, we obtain the following relations:
\begin{align}
 - C\, S_{\hat{\cH}} \, \ket{\hat{F}} 
 = - \sum_p \frac{1}{p!}\, \hat{F}_{m_1\cdots m_p}\, C\, S_{\hat{\cH}}\gamma^{m_1\cdots m_p}\ket{0} 
 = \sum_p (-1)^{\frac{p(p+1)}{2}+1}\, \ket{*\hat{F}_p} \,, 
\end{align}
and the self-duality relation \eqref{eq:self-duality} becomes
\begin{align}
 * \hat{F}_p = (-1)^{\frac{p(p+1)}{2}+1} \hat{F}_{10-p} \,,\qquad 
 \hat{F}_p = (-1)^{\frac{p(p-1)}{2}} * \hat{F}_{10-p} \,. 
\label{eq:self-duality-form}
\end{align}
From this relation and the Bianchi identity for $\hat{F}$,
\begin{align}
 \rmd \hat{F} + H_3\wedge \hat{F} = 0 \,,
\end{align}
the equation of motion becomes
\begin{align}
 \rmd * \hat{F} - H_3\wedge * \hat{F} =0 \,,
\end{align}
where, compared to the Bianchi identity, the sign in front of $H_3$ is flipped, 
according to the sign in \eqref{eq:self-duality-form}. 

\medskip

From \eqref{eq:inner-product}, the action \eqref{eq:Lagrangian-full} becomes
\begin{align}
 \cL =\sqrt{\abs{\CG}} \biggl[\Exp{-2\Phi} \Bigl(R + 4\,D^m \partial_m \Phi - 4\,
\abs{\partial \Phi}^2 - \frac{1}{2}\,\abs{H_3}^2\Bigr) - \frac{1}{4} \sum_p \abs{\hat{F}_p}^2\biggr] \,. 
\end{align}
In order to evaluate the equations of motion \eqref{eq:EOM-DFT}, 
let us recall that $\cS_{MN}$ takes the form \cite{Sakatani:2016fvh}
\begin{align}
\begin{split}
 (\cS_{MN})
 &=\begin{pmatrix}
 2\,\CG_{(m|k}\, s^{[kl]}\, B_{l|n)} - s_{(mn)} 
 - B_{mk}\, s^{(kl)}\, B_{ln} & B_{mk}\,s^{(kn)} - \CG_{mk}\, s^{[kn]} \\
 s^{[mk]}\, \CG_{kn} -s^{(mk)}\, B_{km} & s^{(mn)}
 \end{pmatrix} \,,
\\
 s_{(mn)} &\equiv R_{mn}-\frac{1}{4}\,H_{mpq}\,H_n{}^{pq} + 2 D_m \partial_n \Phi \,,
\qquad
 s_{[mn]} \equiv - \frac{1}{2}\,D^k H_{kmn} + \partial_k\Phi\,H^k{}_{mn} \,. 
\end{split}
\end{align}
In fact, $\cE_{MN}$ also takes a similar form. 
From the rewriting
\begin{align}
 \cE_{MN} = \frac{1}{4}\Exp{2d}\,\bra{\hat{F}}\,\bigl[(\gamma_K)^\rmT\,S_{\hat{\cH}}\,\gamma_L 
-\tfrac{1}{2}\,\hat{\cH}_{KL}\,S_{\hat{\cH}}\bigr]\,\ket{\hat{F}} \,(\Lambda_{-B^\rmT})^K{}_M
\,(\Lambda_{-B})^L{}_N\,,
\end{align}
it is straightforward to derive 
\begin{align}
\begin{split}
 (\cE_{MN})&= \begin{pmatrix}
 -2\,\CG_{(m|k}\, \cK^{kl}\, B_{l|n)} + T_{mn} 
 + B_{mk}\, T^{kl}\, B_{ln} & -B_{mk}\,T^{kn} + \CG_{mk}\, \cK^{kn} \\
 -\cK^{mk}\, \CG_{kn} + T^{mk}\, B_{km} & -T^{mn}
 \end{pmatrix} \,,
\\
 T_{mn} &\equiv \frac{1}{4} \Exp{2\Phi}\, \sum_p \Bigl[ \frac{1}{(p-1)!}\, 
\hat{F}_{(m}{}^{k_1\cdots k_{p-1}} \hat{F}_{n) k_1\cdots k_{p-1}} 
- \frac{1}{2}\, \CG_{mn}\,\abs{\hat{F}_p}^2 \Bigr] \,,
\\
 \cK_{mn}&\equiv \frac{1}{4}\Exp{2\Phi}\, \sum_p \frac{1}{(p-2)!}\, 
\hat{F}_{k_1\cdots k_{p-2}}\, \hat{F}_{mn}{}^{k_1\cdots k_{p-2}} \,. 
\end{split}
\end{align}
Thus, the equations of motion \eqref{eq:EOM-DFT} are summarized as
\begin{align}
\begin{alignedat}{2}
 &R_{mn}-\frac{1}{4}\,H_{mpq}\,H_n{}^{pq} + 2 D_m \partial_n \Phi = T_{mn}\,, \quad&
 &R + 4\,D^m \partial_m \Phi - 4\,\abs{\partial \Phi}^2 - \frac{1}{2}\,\abs{H_3}^2 = 0 \,,
\\
 &- \frac{1}{2}\,D^k H_{kmn} + \partial_k\Phi\,H^k{}_{mn} = \cK_{mn} \,,\quad& 
 &\rmd * \hat{F}_p - H_3\wedge * \hat{F}_{p+2} =0 \,,
\end{alignedat}
\end{align}
where $p$ is even/odd for type IIA/IIB supergravity.

\subsection{Generalized type IIA/IIB equations}

As discussed in Sect.~\ref{sec:mDFT-is-DFT}, 
the equations of motion for the (m)DFT can be obtained 
by introducing a linear $\tilde{x}_m$ dependence 
into the DFT dilaton or the conventional dilaton,
\begin{align}
 d ~~\to~~ d_*\equiv d + I^m\,\tilde{x}_m \,,\qquad \Phi ~~\to~~ \Phi_* \equiv \Phi + I^m\,\tilde{x}_m \,. 
\end{align}
According to this replacement, the equations of motion for the NS--NS sector 
are modified \cite{Sakatani:2016fvh}. 
On the other hand, because there is no dilaton dependence in the equations of motion for the R--R sector, 
one may deduce that the R--R sector should not be modified. 
However, as we see in various examples \cite{SSY2}, the R--R fields $\ket{A}$ or $\ket{F}$ 
already include a non-linear dual-coordinate dependence through the dilaton $\Phi_*$,
\begin{align}
 \ket{A} = \Exp{-\Phi_*} \ket{\cA} \,,\qquad \ket{F} = \Exp{-\Phi_*} \ket{\cF} \,, 
\end{align}
where $\cA$ and $\cF$, to be called the $\Phi_*$-untwisted fields, are supposed 
to be independent of the dual coordinates. 
In fact, these rescaled fields, $\cA$ and $\cF$, appear more naturally 
in the vielbein formulations of the DFT discussed in 
\cite{Jeon:2012kd} and \cite{Geissbuhler:2011mx,Geissbuhler:2013uka}.

\medskip

In terms of the $\Phi_*$-untwisted fields, we obtain the relation
\begin{align}
 \ket{\cF} &= \Exp{\Phi_*}\sla{\partial}(\Exp{-\Phi_*} \ket{\cA}) 
 = \sla{\partial} \ket{\cA} - (\sla{\partial}\Phi_*)\,\ket{\cA} 
 = \gamma^m\,(\partial_m - \partial_m\Phi)\,\ket{\cA} - I^m \,\gamma_m\,\ket{\cA} 
\nn\\
 &= \ket{\rmd\cA - \rmd\Phi\wedge \cA - \iota_I \cA} \,. 
\end{align}
The Bianchi identity is also modified in a similar manner,
\begin{align}
 0 = \Exp{\Phi_*}\sla{\partial}(\Exp{-\Phi_*} \ket{\cF}) 
   = \ket{\rmd\cF - \rmd\Phi\wedge \cF - \iota_I \cF} \,.
\end{align}
More explicitly, we obtain
\begin{align}
 \cF_{p+1} =\rmd\cA_p - \rmd\Phi\wedge \cA_p - \iota_I \cA_{p+2}\,,\qquad 
 \rmd\cF_p - \rmd\Phi\wedge \cF_p - \iota_I \cF_{p+2} =0 \,.
\end{align}

\medskip

If we further introduce the $(\Phi_*,B)$-untwisted fields%
\footnote{The flat components of $\hat{\cC}$ correspond to the fundamental fields in 
\cite{Geissbuhler:2011mx,Geissbuhler:2013uka} 
by taking the conventional parameterization of the generalized vielbein 
in terms of the vielbein for $\CG_{mn}$ and $B_{mn}$.} defined as 
\begin{align}
 \hat{\cF} \equiv \Exp{-B_2\wedge} \cF \,,\qquad \hat{\cC} \equiv \Exp{-B_2\wedge} \cA \,, 
\end{align}
the relation between the potential and the field strength is represented by 
\begin{align}
 \hat{\cF} &= \Exp{-B_2\wedge} \bigl[\rmd (\Exp{B_2\wedge}\hat{\cC}) 
- \rmd\Phi\wedge (\Exp{B_2\wedge}\hat{\cC}) - \iota_I (\Exp{B_2\wedge}\hat{\cC})\bigr] 
\nn\\
 &= \rmd \hat{\cC} - Z\wedge \hat{\cC} - \iota_I \hat{\cC} + H_3\wedge \hat{\cC} \qquad (Z\equiv \rmd\Phi +\iota_I B_2)\,. 
\end{align}
Namely, the $(p+1)$-form field strength is given by
\begin{align}
 \hat{\cF}_{p+1} = \rmd \hat{\cC}_p - Z \wedge \hat{\cC}_p - \iota_I \hat{\cC}_{p+2} + H_3 \wedge \hat{\cC}_{p-2} \,. 
\end{align}
The Bianchi identity becomes
\begin{align}
 \rmd \hat{\cF}_p - Z \wedge \hat{\cF}_p - \iota_I \hat{\cF}_{p+2} + H_3 \wedge \hat{\cF}_{p-2} = 0\,.
\end{align}
The gauge transformation of the R--R potential is expressed as 
\begin{align}
 \delta_\lambda \hat{\cC} = \rmd \hat{\lambda} - Z \wedge \hat{\lambda} - \iota_I \hat{\lambda} 
+ H_3 \wedge \hat{\lambda} \,,
\end{align}
and the invariance of the field strength requires the nilpotency
\begin{align}
 0=(\rmd - Z \wedge - \iota_I + H_3 \wedge)^2 \hat{\lambda} 
= - \Lie_I \hat{\lambda} - \bigl(\rmd Z + \iota_I H_3 - \iota_I Z\bigr) \wedge \hat{\lambda} \,. 
\label{eq:nilpotency}
\end{align}
But this is indeed satisfied from the strong constraint 
\eqref{eq:condition-section} and $\Lie_I \lambda=0$\,. 

\medskip

The equations of motion become (see \cite{Sakatani:2016fvh} for the modification of the NS--NS sector)
\begin{align}
 &R_{mn}-\frac{1}{4}\,H_{mpq}\,H_n{}^{pq} + 2 D_m \partial_n \Phi + D_m U_n +D_n U_m = T_{mn} \,,
\nn\\
 &R + 4\,D^m \partial_m \Phi - 4\,\abs{\partial \Phi}^2 - \frac{1}{2}\,\abs{H_3}^2 
 - 4\,\bigl(I^m I_m+U^m U_m + 2\,U^m\,\partial_m \Phi - D_m U^m\bigr) =0 \,,
\nn\\
 &-\frac{1}{2}\,D^k H_{kmn} + \partial_k\Phi\,H^k{}_{mn} + U^k\,H_{kmn} + D_m I_n - D_n I_m = \cK_{mn} \,,
\label{eq:GSE}
\\
 &\rmd *\hat{\cF}_p - Z \wedge * \hat{\cF}_p -\iota_I * \hat{\cF}_{p-2} 
- H_3\wedge * \hat{\cF}_{p+2} =0 \,,
\nn
\end{align}
where we have introduced the following quantities: 
\begin{align}
\begin{split}
 T_{mn} &= \frac{1}{4} \sum_p \Bigl[ \frac{1}{(p-1)!}\, 
\hat{\cF}_{(m}{}^{k_1\cdots k_{p-1}} \hat{\cF}_{n) k_1\cdots k_{p-1}} - \frac{1}{2}\, 
\CG_{mn}\,\abs{\hat{\cF}_p}^2 \Bigr] \,,
\\
 \cK_{mn}&= \frac{1}{4} \sum_p \frac{1}{(p-2)!}\, \hat{\cF}_{k_1\cdots k_{p-2}}\, 
\hat{\cF}_{mn}{}^{k_1\cdots k_{p-2}} \,. 
\end{split}
\end{align}
For the type IIB case, these equations are nothing but 
the generalized type IIB equations (in the democratic form).

\paragraph*{Redefinitions of the Ramond--Ramond fields\\}

Our fundamental fields are $(A,\,F)$. However, depending on the situation, 
it may be more convenient to introduce various untwisted quantities. 
For completeness, let us introduce the following redefinitions of the R--R fields:
\begin{align}
\begin{split}
 (\check{C},\,\check{F}) \quad \xleftarrow[\quad \beta\text{-untwist}\quad]{\Exp{\beta\vee}}\quad 
 (A,&\,\,F) \quad \xrightarrow[\quad B\text{-untwist}\quad]{\Exp{-B_2\wedge}}\quad 
 (\hat{C},\,\hat{F})
\\
 \text{\scriptsize$\Exp{I^m \tilde{x}_m}$ }~&\!\!\!\Big\downarrow\text{\scriptsize$I$-untwist}
\\
 (\check{\sfC},\,\check{\sfF}) \quad \xleftarrow[\quad \beta\text{-untwist}\quad]{\Exp{\beta\vee}}\quad 
 (\sfA,&\,\,\sfF) \quad \xrightarrow[\quad B\text{-untwist}\quad]{\Exp{-B_2\wedge}}\quad 
 (\hat{\sfC},\,\hat{\sfF})
\\
 \text{\scriptsize$\Exp{\Phi}$ }~&\!\!\!\Big\downarrow\text{\scriptsize$\Phi$-untwist}
\\
 (\check{\cC},\,\check{\cF}) \quad \xleftarrow[\quad \beta\text{-untwist}\quad]{\Exp{\beta\vee}}\quad 
 (\cA,&\,\,\cF) \quad \xrightarrow[\quad B\text{-untwist}\quad]{\Exp{-B_2\wedge}}\quad 
 (\hat{\cC},\,\hat{\cF})
\end{split}
\end{align}
For example, the $(I,\,B)$-untwisted fields are defined as 
\begin{align}
 \hat{\sfC} \equiv \Exp{I^m\,\tilde{x}_m} \Exp{-B_2\wedge} A
 = \Exp{-\Phi}\cC\,,\qquad \hat{\sfF} \equiv 
\Exp{I^m\,\tilde{x}_m} \Exp{-B_2\wedge} F = \Exp{-\Phi}\,\hat{\cF} \,,
\end{align}
the potentials and the field strengths are related through 
\begin{align}
 \hat{\sfF}_{p+1} = \rmd \hat{\sfC}_p - U \wedge \hat{\sfC}_p - \iota_I \hat{\sfC}_{p+2} 
+ H_3 \wedge \hat{\sfC}_{p-2} \,,
\label{eq:sfF-sfC}
\end{align}
and the Bianchi identities become
\begin{align}
 \rmd \hat{\sfF}_p - U \wedge \hat{\sfF}_p - \iota_I \hat{\sfF}_{p+2} + H_3 \wedge \hat{\sfF}_{p-2} = 0 \,.
\end{align}

\medskip

In particular, when we study non-geometric $T$-folds, it is more convenient 
to define the $(I,\,\beta)$-untwisted fields \cite{SSY2},%
\footnote{A definition of $\beta$-untwisted R--R fields is discussed in \cite{Andriot:2014qla}.}
\begin{align}
 \ket{\check{\sfC}} \equiv \Exp{I^m\,\tilde{x}_m} \Exp{\bbeta}\, \ket{A} \,,\qquad 
\ket{\check{\sfF}} \equiv \Exp{I^m\,\tilde{x}_m} \Exp{\bbeta}\, \ket{F}
\end{align}
or
\begin{align}
 \check{\sfC} \equiv \Exp{I^m\,\tilde{x}_m} \Exp{\beta\vee} A \,,\qquad 
\check{\sfF} \equiv \Exp{I^m\,\tilde{x}_m} \Exp{\beta\vee} F \,,
\end{align}
where we have defined an operation $\beta\vee$ that acts on a $p$-form $\alpha_p$ as
\begin{align}
 \beta\vee \alpha_p \equiv \frac{1}{2} \,\beta^{mn}\,\iota_m\iota_n \alpha_p \,. 
\end{align}
For the $\beta$-untwisted quantities, the natural metric is the metric $\OG_{mn}$ 
introduced in \eqref{eq:non-geometric-parameterization}.
For example, the self-duality relation \eqref{eq:self-duality} takes the form
\begin{align}
 \star \check{F}_p = (-1)^{\frac{p(p+1)}{2}+1} \check{F}_{10-p} \,,\qquad 
 \check{F}_p = (-1)^{\frac{p(p-1)}{2}} \star \check{F}_{10-p} \,,
\end{align}
and the same relations hold for $\check{\cF}$ and $\check{\sfF}$\,, 
where $\star$ is the Hodge star operator associated with $\OG_{mn}$\,. 
Also, if there is no modification $I^m$, the Lagrangian for the R--R sector becomes
\begin{align}
 \cL_{\text{R--R}} = - \frac{1}{4}\sqrt{\abs{\OG}}\, \sum_p\frac{1}{p!}\,\OG^{m_1n_1}\cdots 
\OG^{m_pn_p}\,\check{F}_{m_1\cdots m_p}\,\check{F}_{n_1\cdots n_p} \,. 
\end{align}

\subsection{$T$-duality transformation rules}

In this subsection, we present the $T$-duality transformation rule 
in a coordinate system in which \eqref{eq:X-gauge} is realized. 
If the fields $(\cH_{MN},\,d,\,A_p)$ are independent of a coordinate $z$, 
the following $T$-duality transformation along the $z$-direction maps 
a solution of the (m)DFT to another one of the (m)DFT:
\begin{align}
\begin{split}
 &\cH_{MN}\to \cH'_{MN}=(\Lambda^\rmT\,\cH\,\Lambda)_{MN} \,,\qquad 
 \Lambda\equiv (\Lambda^M{}_N) \equiv \begin{pmatrix} \delta^m_n - \delta^m_z\,\delta_n^z 
& \delta^m_z\,\delta^n_z \\ \delta_m^z\,\delta_n^z 
& \delta_m^n - \delta_m^z\,\delta^n_z \end{pmatrix}\,, 
\\
 &\ket{A}_{\text{IIA}} \to \ket{\cA'}_{\text{IIB}} = (\gamma_z -\gamma^z)\,\ket{\cA}_{\text{IIA}} \,,\qquad 
 \ket{A}_{\text{IIB}} \to \ket{\cA'}_{\text{IIA}} = (\gamma^z -\gamma_z)\,\ket{\cA}_{\text{IIB}} \,, 
\\
 &d\to d'= d + I^z\,z\,, \qquad I^z \to I'^z=0\,,\qquad I^i \to I'^i=I^i \,.
\end{split}
\end{align}
Here, $x^i$ is an arbitrary coordinate other than $z$\,. 
In the component expression, the above rule is represented by the following map:
\begin{align}
\begin{split}
 &\CG'_{ij} = \CG_{ij} - \frac{\CG_{iz}\, \CG_{jz}- B_{iz}\, B_{jz}}{\CG_{zz}}\,,\qquad 
 \CG'_{iz} = \frac{B_{iz}}{\CG_{zz}}\,,\qquad 
 \CG'_{zz}=\frac{1}{\CG_{zz}}\,,
\\
 &B'_{ij} = B_{ij} - \frac{B_{iz}\, \CG_{jz}- \CG_{iz}\, B_{jz}}{\CG_{zz}}\,,\qquad 
 B'_{iz} = \frac{\CG_{iz}}{\CG_{zz}} \,, 
\\
 &\Phi' =\Phi + \frac{1}{4}\,\ln\Bigl\lvert \frac{\det \CG'_{mn}}{\det \CG_{mn}} \Bigr\rvert 
+ I^z z \,,\qquad 
 I'^i = I^i\,,\qquad I'^z = 0 \,,
\\
 &\cA'_{i_1\cdots i_{p-1}z} = \cA_{i_1\cdots i_{p-1}} \,,
\qquad
 \cA'_{i_1\cdots i_p} = \cA_{i_1\cdots i_pz} \,. 
\end{split}
\end{align}
For the other R--R fields, the transformation rules are given by %for example, 
\begin{align}
\begin{split}
 &\sfA'_{i_1\cdots i_{p-1}z} = \Exp{-I^z z} \sfA_{i_1\cdots i_{p-1}} \,,
\qquad
 \sfA'_{i_1\cdots i_p} = \Exp{-I^z z} \sfA_{i_1\cdots i_pz} \,,
\\
 &\hat{\cC}'_{i_1\cdots i_{p-1}z} = \hat{\cC}_{i_1\cdots i_{p-1}} 
- (p-1)\,\frac{\hat{\cC}_{[i_1\cdots i_{p-2}|z|}\, \CG_{i_{p-1}]z}}{\CG_{zz}} \,,
\\
 &\hat{\cC}'_{i_1\cdots i_p} = \hat{\cC}_{i_1\cdots i_pz} + p\, 
\hat{\cC}_{[i_1\cdots i_{p-1}}\, B_{i_p]z} + p\,(p-1)\,\frac{\hat{\cC}_{[i_1\cdots i_{p-2}|z|}\, 
B_{i_{p-1}|z|}\, \CG_{i_p]z}}{\CG_{zz}} \,. 
\end{split}
\end{align}

\section{Scale invariance and Weyl invariance}
\label{sec:Weyl}

Let us recall the conventional bosonic string sigma model in the critical dimension $D=26$,%
\footnote{In the following discussion, we will not show the ghost sector explicitly.}
\begin{align}
 S =-\frac{1}{4\pi\alpha'} \int \rmd^2\sigma \sqrt{-\gamma}\,\bigl(\CG_{mn}\,\gamma^{ab} - B_{mn}\,\varepsilon^{ab}\bigr)\, \partial_a X^m\, \partial_b X^n \,,
\end{align}
where $a,b=\tau,\sigma$, $\varepsilon^{\tau\sigma}=+1/\sqrt{-\gamma}$, and $\varepsilon_{\tau\sigma}=-\sqrt{-\gamma}$\,. 
This system is scale invariant (or UV finite) at the one-loop level \cite{Hull:1985rc}, if the beta functions
\begin{align}
 \beta^{\CG}_{mn} = \alpha'\,\Bigl(R_{mn}-\frac{1}{4}\,H_{mpq}\,H_n{}^{pq}\Bigr) \,,\qquad
 \beta^{B}_{mn} = \alpha'\,\Bigl(- \frac{1}{2}\,D^k H_{kmn}\Bigr) \,,
\end{align}
take the following forms:\footnote{In terms of the DFT, these equations can neatly be
summarized as $\cS_{MN}=\gLie_\bY \cH_{MN}$ \cite{Sakatani:2016fvh}, 
where $(\bY^M)=(U^m,\ I_m + B_{mn}\,U^n)$ is an arbitrary generalized vector and 
$Z_m=\partial_m\Phi + U_m$\,.}
\begin{align}
 \beta^{\CG}_{mn} = - 2\,\alpha'\,D_{(m} Z_{n)}\,,\qquad 
\beta^{B}_{mn} = - \alpha'\,\bigl(Z^k\,H_{kmn} + 2\,D_{[m} I_{n]}\bigr) \,,
\label{eq:finiteness}
\end{align}
where $I^m$ and $Z_m$ are arbitrary vectors. 
The condition for Weyl invariance is more restrictive, and it is satisfied 
only when $I^m$ and $Z_m$ have the following forms: 
\begin{align}
 I^m = 0 \,, \qquad Z_m = \partial_m\Phi \,. 
\label{eq:Weyl-condition}
\end{align}
This is because the trace of the energy--momentum tensor takes the form
\begin{align}
 2\alpha'\,\langle T\rangle = \bigl(\beta^{\CG}_{mn}\,\gamma^{ab} - \beta^{B}_{mn}\, \varepsilon^{ab}\bigr)\, \partial_a X^m\, \partial_b X^n \,,
\label{eq:trace-anomaly}
\end{align}
and the non-vanishing $\beta$-function can be canceled by adding the Fradkin--Tseytlin term,
\begin{align}
 S_{\text{FT}} = \frac{1}{4\pi}\int \rmd^2\sigma \, \sqrt{-\gamma}\,R^{(\gamma)}\,\Phi\,, 
\end{align}
only when \eqref{eq:Weyl-condition} is satisfied. 
Indeed, $S_{\text{FT}}$ gives a contribution to $\langle T\rangle$ as
\begin{align}
 \frac{4\pi}{\sqrt{-\gamma}}\,\gamma^{ab}\,\frac{\delta S_{\text{FT}}}{\delta \gamma^{ab}}
 = \Bigl(D_m D_n \Phi\,\gamma^{ab} - \frac{1}{2}\,\partial_k\Phi\,H^k{}_{mn}\,\varepsilon^{ab}\Bigr)\, 
\partial_a X^m\,\partial_b X^n + D^m \Phi\,\frac{2\pi\alpha'}{\sqrt{-\gamma}}\frac{\delta S}{\delta X^m} \,,
\end{align}
and by using the equations of motion for $X^m$,
\begin{align}
 0 = \frac{2\pi\alpha'}{\sqrt{-\gamma}}\frac{\delta S}{\delta X^m} 
 = G_{mn}\,\biggl[\cD^a \partial_a X^n + \Bigl(\gamma^n_{kl}\,\gamma^{ab} + \frac{1}{2}\,H^n{}_{kl}\,\varepsilon^{ab}\Bigr)\,\partial_a X^k\,\partial_b X^l\biggr] \,,
\end{align}
where the covariant derivative $\cD_a$ is associated with the intrinsic metric $\gamma_{ab}$ and $\gamma^n_{kl}$ is the Christoffel symbol associated with $G_{mn}$, 
this contribution to $\langle T\rangle$ precisely cancels the Weyl anomaly \eqref{eq:trace-anomaly} 
for \eqref{eq:Weyl-condition}. 
For general $I^m$ and $Z_m$, we cannot find a suitable local counterterm 
and the Weyl anomaly cannot be canceled. 

\medskip 

In the following, by considering the doubled spacetime, 
we show that the Fradkin--Tseytlin term can completely cancel the Weyl anomaly 
if $I^m$ and $Z_m$ satisfy the conditions \eqref{eq:GSE-conditions} and \eqref{eq:dilaton-isometry}. 
When these conditions are satisfied, $I^m$ and $Z_m$ can be replaced 
by the (m)DFT dilaton $d_*$ or $\Phi_*$, which has dual-coordinate dependence. 
In order to treat the dual coordinates, we consider the DSM in which 
the number of the embedding functions is doubled: $(\bbX^M)=(X^m,\,\tilde{X}_m)$\,. 
For convenience, we choose the coordinates
\begin{align}
 (x^m) = (x^\mu,\,y^i) \quad (\mu=0,\dotsc,D-N-1;\ i=1,\dotsc,N) 
\end{align}
on the target space such that the background fields $(\CG_{mn},\,B_{mn},\,\Phi)$ 
depend only on $x^\mu$, and the modified dilaton $\Phi_*$ has the form $\Phi_*=\Phi +I^i\,\tilde{y}_i$\,. 
Then, the essence of our argument is that the contribution of the Fradkin--Tseytlin term (with $\Phi$ replaced by $\Phi_*$) to the trace $\langle T\rangle$ can be written as
\begin{align}
 &\frac{4\pi}{\sqrt{-\gamma}}\,\gamma^{ab}\,\frac{\delta S_{\text{FT}}}{\delta \gamma^{ab}}
 = \cD^a \bigl[\partial_M\Phi_*(\bbX)\,\partial_a \bbX^M \bigr] 
 = \cD^a \bigl[\partial_\mu\Phi(X)\, \partial_a X^\mu + I^i\,\partial_a \tilde{Y}_i \bigr] 
\nn\\
 &= \cD^a \bigl(Z_m\,\partial_a X^m \bigr) - \partial_{[m} I_{n]} \,\varepsilon^{ab}\,\partial_a X^m\,\partial_b X^n 
  + \cD^a \bigl[I^i\,(\partial_a \tilde{Y}_i - \CG_{in}\, \varepsilon^b{}_a\,\partial_b X^n - B_{in}\,\partial_a X^n) \bigr] 
\nn\\
 &= \frac{1}{2}\, \bigl[2\,D_{(m} Z_{n)}\,\gamma^{ab} - (Z^k\,H_{kmn}+2\,\partial_{[m} I_{n]})\,
\varepsilon^{ab} \bigr]\,\partial_a X^m \, \partial_b X^n 
\nn\\
 &\quad +Z^m \,\frac{2\pi\alpha'}{\sqrt{-\gamma}}\frac{\delta S}{\delta X^m}
 + I^i\,\cD^a (\partial_a \tilde{Y}_i - \CG_{in}\, \varepsilon^b{}_a\,\partial_b X^n - B_{in}\,\partial_a X^n) \,. 
\label{eq:Weyl-inv}
\end{align}
By using the equations of motion for $X^m$ and the self-duality relations
\begin{align}
 \partial_a \tilde{Y}_i - \CG_{in}\, \varepsilon^b{}_a\,\partial_b X^n - B_{in}\,\partial_a X^n =0 \,,
\label{eq:self-duality-components}
\end{align}
which are also obtained as the equations of motion of the DSM, as we will explain below, the contribution from the Fradkin--Tseytlin term can completely cancel the Weyl anomaly \eqref{eq:trace-anomaly}. 

\medskip 

In order to explain the self-duality relation, let us consider Hull's double sigma model,
\begin{align}
 S = \frac{1}{4\pi\alpha'}\int \Bigl[ \frac{1}{2}\,\cH_{MN}(X)\,\cP^M\wedge *_\gamma 
\cP^N - \bigl(\rmd \tilde{X}_m+C_m\bigr)\wedge \rmd X^m \Bigr] \,,
\end{align}
where we have introduced the quantities 
\begin{align}
 \cP^M(\sigma) \equiv \rmd \bbX^M(\sigma) + C^M(\sigma) \,,\qquad 
(C^M)=\begin{pmatrix} 0 \\ C_m \end{pmatrix}\,,
\end{align}
and the generalized metric $\cH_{MN}(X)$ are supposed to be independent of 
$(Y^I)\equiv (Y^i,\,\tilde{Y}_i)$\,. 
The equations of motion for $C_m$ give rise to 
\begin{align}
 \rmd \tilde{X}_m + C_m = \CG_{mn}\,*_\gamma \rmd X^n + B_{mn}\,\rmd X^n \,,
\end{align}
which is equivalent to the well-known self-duality relation
\begin{align}
 \cP_M = *_\gamma \cH_M{}^N\,\cP_N \,.
\end{align}
Using the above equations of motion for $C_m$ and eliminating the combination 
$\rmd \tilde{X}_m + C_m$ from the above action, 
we obtain the conventional string sigma model action for $X^m$\,,
\begin{align}
 S = \frac{1}{4\pi\alpha'}\int \bigl(\CG_{mn}\,\rmd X^m\wedge *_\gamma \rmd X^n + B_{mn}\,\rmd X^m\wedge\rmd X^n \bigr) \,. 
\end{align}
Thus, the DSM is classically equivalent to the conventional sigma model. 
By combining the equations of motion for $Y^i$ and $C_m$, we can show that 
\begin{align}
 \rmd C_i = \rmd \bigl(\CG_{in}\,*_\gamma \rmd X^n + B_{in}\,\rmd X^n\bigr) = 0\,,
\end{align}
and $C_i$ is a closed form. 
Thus, using the local symmetry
\begin{align}
 \tilde{Y}_i(\sigma) \to \tilde{Y}_i(\sigma) + \alpha_i(\sigma)\,,\qquad 
C_i(\sigma) \to C_i -\rmd \alpha_i(\sigma)\,,
\end{align}
we can (at least locally) set $C_i=0$, and the equations of motion for $C_i$ become
\begin{align}
 \rmd \tilde{Y}_i = \CG_{in}\,*_\gamma \rmd X^n + B_{in}\,\rmd X^n \,. 
\end{align}
This is nothing but the key relation \eqref{eq:self-duality-components}. 

\medskip 

The one-loop $\beta$-function of the DSM is computed in 
\cite{Berman:2007xn,Berman:2007yf,Copland:2011yh} by using the background field method, 
and the result is consistent with the conventional string sigma model. 
There, in order to cancel the Weyl anomaly, the $T$-duality-invariant Fradkin--Tseytlin term \cite{Hull:2006va},%
\footnote{This reduces to the conventional one after integrating over the auxiliary fields $C_m$ \cite{Hull:2006va}.}
\begin{align}
 S_{\text{FT}} = \frac{1}{8\pi}\int \rmd^2\sigma \, \sqrt{-\gamma}\,R^{(\gamma)}\,d(X)\,,
\end{align}
has been introduced, though the dilaton is supposed to be independent of the dual coordinates 
and the equations of motion of the (m)DFT have not been reproduced. 
By replacing $d(X)$ with $d_*(X,\,\tilde{Y})$\,, Weyl invariance follows from the same 
calculation as \eqref{eq:Weyl-inv}. 
Therefore, the bosonic string sigma model is Weyl invariant, as long as the background fields 
satisfy the GSE. 
When we consider the type II Green--Schwarz superstring theories, the $\beta$-functions receive additional corrections 
coming from the R--R field strength [i.e., $T_{mn}$ and $\cK_{mn}$ in \eqref{eq:GSE}] 
\cite{Arutyunov:2015mqj}. 
However, there is no explicit $I^m$ and $Z_m$ dependence in $T_{mn}$ and $\cK_{mn}$, 
hence the (modified) Fradkin--Tseytlin term is enough for the cancelation of the Weyl anomaly 
for the NS--NS sector of the general backgrounds of the GSE. 
As for the $\beta$-functions for the R--R fields, we still have much to do. 

\medskip 

In the literature \cite{Hull:1985rc,Tseytlin:1986tt,Shore:1986hk,Tseytlin:1986ws}, 
the difference between scale invariance (or finiteness) and Weyl invariance has been studied, 
but our result has alleviated the gap. 
The only thing remaining is that the scale invariance does not require any condition for $I^m$ and $Z_m$, 
while the Weyl invariance is realized only when the conditions in \eqref{eq:GSE-conditions} and \eqref{eq:dilaton-isometry} are satisfied. 
We expect that this small difference is intrinsic to the bosonic string, and 
when we consider the type II Green--Schwarz superstring theory 
the scale invariance may require the same conditions \eqref{eq:GSE-conditions} and \eqref{eq:dilaton-isometry}, 
which are also required from the kappa symmetry of the superstring theory. 

\medskip 

Before closing this section, let us comment on the central charge identity \cite{Callan:1985ia,Curci:1986hi}, 
namely the constancy of $\cS$. 
As discussed in \cite{Sakatani:2016fvh}, one can show the identity only 
from the differential Bianchi identity and the equations of motion for the generalized metric, $\cS_{MN}=0$,
\begin{align}
 \partial_M \cS = 2\,\cH_{MN}\,\nabla_K \cS^{KN} = 0\,. 
\end{align}
In \cite{Sakatani:2016fvh}, the differential Bianchi identity has not been proven 
in the presence of $\bX^M$, and it has not been clear 
whether $\bS_{MN}=0$ can generally show the central charge identity. 
However, as we have found that the mDFT is just the conventional DFT, 
the differential Bianchi identity and $\bS_{MN}=0$ indicate that $\bS$ is constant. 

\section{Conclusion and discussion}
\label{sec:discussion}

In this paper, the bosonic part of the GSE is completely reproduced from the (m)DFT. 
When all of the fields are independent of the dual coordinates, 
the equations of motion of the DFT lead to the conventional supergravity equations, 
while when the dilaton has a linear dual-coordinate dependence, the GSE are reproduced. 
The type IIA supergravity equations of motion have been presented in the same manner 
as the type IIB equations and the $T$-duality transformation rules between the type IIA and the type IIB have been provided. 
The discrepancy between the scale invariance and the Weyl invariance of the bosonic string theory has been resolved 
by introducing dual-coordinate dependence into the Fradkin--Tseytlin term. 
The existence of the doubled space is indispensable to our discussion. 

\medskip 

In this paper, we have considered a $T$-duality covariant language of the DFT for simplicity, 
but the GSE can also be derived \cite{Baguet:2016prz} from a $U$-duality covariant generalization, 
the exceptional field theory (EFT) \cite{West:2001as,West:2003fc,West:2004st,Hillmann:2009ci,Berman:2010is,Berman:2011cg,Berman:2011jh,Berman:2012vc,Hohm:2013pua,Hohm:2013vpa,Hohm:2013uia,Aldazabal:2013via,Hohm:2014fxa}. 
When we describe the type IIB theory in the $E_{d(d)}$ EFT (which was initially done in \cite{Hohm:2013pua,Blair:2013gqa}), following the convention of \cite{Sakatani:2017nfr}, 
the generalized coordinates are introduced as 
\begin{align}
 (x^\mu,\ \sfx^\sfM)=(x^\mu,\ \sfx^\sfm,\ \sfy_\sfm^\alpha,\ \sfy_{\sfm_1\sfm_2\sfm_3},\ 
 \sfy_{\sfm_1\cdots \sfm_5}^\alpha,\ \sfy_{\sfm_1\cdots \sfm_6,\,\sfm},\cdots) \,,
\end{align}
where the ellipsis is not necessary for $d\leq 7$\,. 
Here, $x^\mu$ ($\mu=0,10-d$) are the coordinates on the non-compact directions, 
$\sfx^\sfm$ ($\sfm=1,\dotsc,d-1$) are the conventional coordinates on the torus $T^{d-1}$, 
and the other coordinates are winding coordinates associated with various type IIB branes 
(see \cite{Sakatani:2017nfr} for more detail). 
The index $\alpha=1,2$ is for the fundamental representation of the $\mathrm{SL}(2)$ $S$-duality symmetry. 
In the exceptional space, a doubled torus can be seen as a $2(d-1)$-dimensional space with coordinates $(\sfx^\sfm,\,\sfy^1_\sfm)$\,. 
For convenience, let us consider the decomposition
\begin{align}
 (\sfx^\sfm)=(\sfx^\sfa,\,\sfx^\sfi) \quad (\sfa=1,\dotsc,d-N-1;\ \sfi=d-N,\dotsc,d-1)\,. 
\end{align}
Then, if the dilaton depends on $(x^\mu,\,\sfx^\sfa)$ and also has 
a linear coordinate dependence on $\sfy^1_\sfi$, and all of the other fields are functions of 
$(x^\mu,\,\sfx^\sfa)$, the GSE can be reproduced from the equations of motion of the EFT. 
In the case of $d=6$ and $N=1$, where $\sfy_{*+} \equiv \sfy^1_\sfi$, 
the coordinates $(\sfx^\sfa,\,\sfy_{*+})$ are precisely the coordinates used in 
\cite{Baguet:2016prz}, which can be identified with the conventional coordinates in the type IIA theory 
(from the linear map of \cite{Sakatani:2017nfr}). 
In this sense, the result of this paper is consistent with that of \cite{Baguet:2016prz}. 
However, the Scherk--Schwarz ansatz employed in \cite{Baguet:2016prz} was not necessary in our approach. 
The dual derivatives of the dilaton generate the extra vector fields, and the GSE 
follows automatically from the equations of motion of the EFT. 
On the other hand, when we consider the generalized type IIA supergravity, 
the generalized coordinates are parameterized as
\begin{align}
 (x^\mu,\ x^I) = (x^\mu,\ x^i\,,\ y_{i_1i_2}\,,\ y_{i_1\cdots i_5}\,,\ y_{i_1\cdots i_7,\,i}\,, \cdots) \,,
\end{align}
and the $\tilde{x}_i$ coordinates of the DFT can be identified with $y_{iz}$ \cite{Sakatani:2017nfr}, 
where $x^z$ is the coordinate on the $M$-theory circle. 
Thus, if the dilaton has linear dependence on $y_{iz}$, the equations of motion of the EFT 
lead to the generalized type IIA supergravity equations. 

\medskip 

Note that the dual coordinate $\sfy^1_\sfi$ in the exceptional space is not invariant under the $S$-duality symmetry. 
The dilaton also is not a singlet under the $S$-duality. 
Thus, the generalized type IIB supergravity is not covariant under the $S$-duality. 
Namely, under the $S$-duality transformation, a solution of the GSE is mapped 
to a configuration which does not satisfy the GSE, although it is still a solution of the EFT. 
It may be interesting to see whether there exists a $U$-duality-covariant generalization of the GSE. 
Of course, we already know that the EFT can describe all such solutions, and the EFT may be enough. 
In addition, from the point of view of the eleven-dimensional supergravity, 
the dilaton is the eleventh component of the metric, and the introduction of 
dual-coordinate dependence into the dilaton breaks the eleven-dimensional general covariance. 
That would be the reason why the ``generalized eleven-dimensional supergravity'' has not been found. 
Again, it is not necessary to look for such a generalized eleven-dimensional theory 
and the EFT is enough, which has a huge gauge symmetry that includes the eleven-dimensional diffeomorphisms. 

\medskip 

Utilizing the result of this paper, we can straightforwardly study a supersymmetric generalization of the GSE in detail. 
The type II supersymmetric DFT has been constructed in \cite{Jeon:2012hp}. 
By taking a non-standard section of the (m)DFT, one can identify how the extra vector fields 
modify the equations of motion and supersymmetry transformations. 
It will be important to complete this analysis and the obtained result should be compared with that of \cite{WT}. 

\medskip 

It is also important to study a supersymmetric generalization of the DSM analysis. 
The GSE can be derived from the requirement of kappa symmetry, 
while Weyl invariance (at quantum level) has not followed directly from kappa symmetry (at a classical level). 
The result obtained here implies the equivalence of kappa symmetry 
and Weyl invariance, and detailed analysis based on a supersymmetric DSM, 
such as \cite{HackettJones:2006bp,Blair:2013noa,Bandos:2015cha,Park:2016sbw}, would be important future work. 

\medskip 

The present work has provided positive evidence that 
superstring theories defined on solutions of the GSE are Weyl invariant. 
We further expect that superstring theories can be consistently defined on arbitrary solutions of the DFT/EFT. 
It will be an interesting future direction to be addressed. 

\subsection*{Acknowledgments}

We are grateful to J.-H.~Park and A.~A.~Tseytlin for insightful discussions, 
and especially to S.~Uehara for collaboration at an earlier stage. 
We would like to acknowledge useful discussions during the workshop ``Geometry, Duality and Strings'' 
at the Yukawa Institute for Theoretical Physics, Kyoto University. 
The work of J.S.\ was supported by the Japan Society for the Promotion of Science (JSPS). 
The work of K.Y.\ was supported by the Supporting Program for Interaction-based Initiative Team Studies (SPIRITS) 
from Kyoto University and by a JSPS Grant-in-Aid for Scientific Research (C) No.\,15K05051. 
This work was also supported in part by the JSPS Japan--Russia Research Cooperative Program 
and the JSPS Japan--Hungary Research Cooperative Program.

\end{document}